\documentclass[11pt]{article}
\usepackage{amsmath,amsfonts,amssymb}

\DeclareMathOperator{\tr}{Tr}

\def\xx{\mathbi{x}}

\def\AA{\mathbi{A}}
\def\uu{\mathbi{u}}
\def\vv{\mathbi{v}}
\def\kk{\mathbi{k}}

\def\pa{\partial}

\def\div{{\rm div}}

\def\curl{{\rm curl}}

\def\hess{{\rm hess}}
\def\pf{{\rm pf}}

\DeclareMathAlphabet{\mathbi}{OML}{cmm}{b}{it} 

\newcommand{\bu}{\mathbi{u}}

\newcommand{\bv}{\mathbi{v}}
\newcommand{\bvt}{\tilde{\mathbi{v}}}
\newcommand{\bw}{\mathbi{\omega}}

\newcommand{\bel}{\begin{equation}\label}
\newcommand{\ee}{\end{equation}}
\newcommand{\beqa}{\begin{eqnarray}\label}
\newcommand{\eeqa}{\end{eqnarray}} 
\newcommand{\eq}{\end{eqnarray}} 
\newcommand{\bc}{\begin{center}} 
\newcommand{\ec}{\end{center}} 
\newcommand{\ben}{\begin{enumerate}}
\newcommand{\een}{\end{enumerate}}

\newcommand{\bi}{\hat{\mathbi{i}}}
\newcommand{\bj}{\hat{\mathbi{j}}}
\newcommand{\bk}{\hat{\mathbi{k}}}
\newcommand{\cD}{\mathcal{D}}
\newcommand\shalf{\ensuremath{{\scriptstyle\frac{1}{2}}}}
\newcommand\squart{\ensuremath{{\scriptstyle\frac{1}{4}}}}
\newcommand\thhalf{\ensuremath{{\scriptstyle\frac{3}{2}}}}
\newcommand{\twohd}{two-and-a-half-dimensional~}

\newcommand{\bx}{\mathbi{x}}

\newtheorem{theorem}{Theorem}
\newtheorem{lemma}{Lemma}
\newtheorem{proposition}{Proposition}

\newcommand{\beq}{\begin{equation}}
\newcommand{\eeq}{\end{equation}}
\newcommand{\drm}{ {\rm d} }
\newcommand{\Drm}{ {\rm D} }

\begin{document}

\title{K\"ahler Geometry and the Navier-Stokes Equations}
\author{I. Roulstone\thanks{Department of Mathematics and Statistics, University of
Surrey, Guildford GU2 7XH, UK. \textit{email}: i.roulstone@surrey.ac.uk}~,
~B. Banos\thanks{Fachbereich Mathematik, Universit\"at Hamburg, Bundesstrasse 55, D-20146 
Hamburg, Germany}~, 
J. D. Gibbon\thanks{Department of Mathematics, Imperial College London, London SW7 2AZ, UK.}
~~and V. N. Roubtsov\thanks{D\'epartment de Math\'ematiques, Universit\'e d'Angers, 2 Blvd Lavoisier,
49045 Angers, France; \& Theory Division, ITEP, 25, Bol. Tcheremushkinskaya, Moscow 117259, Russia}}
\date{5th September 2005}

\maketitle
\vspace{-8mm}
\begin{abstract}
We study the Navier-Stokes and Euler equations of incompressible hydrodynamics in two and three 
spatial dimensions and show 
how the constraint of incompressiblility leads to equations of Monge--Amp\`ere type for the stream 
function, when the Laplacian of the pressure is known. In two dimensions a K\"ahler geometry is 
described, which is associated with the Monge--Amp\`ere problem. This K\"ahler structure is then 
generalised to `two-and-a-half dimensional' flows, of which Burgers' vortex is one example. In 
three dimensions, we show how a generalized Calabi--Yau structure emerges in a special case. 
\end{abstract}

\section{Equations for an Incompressible Fluid}\label{S1}

Flow visualization methods, allied to large-scale computations of the three-dimensional
incompressible Navier-Stokes equations, vividly illustrate the fact that vorticity has 
a tendency to accumulate on `thin sets' whose morphology is characterized by quasi one-dimensional 
tubes or filaments and quasi two-dimensional sheets. This description is in itself approximate as 
these thin structures undergo dramatic morphological changes in time and space. The topology is 
highly complicated; sheets tend to roll-up into tube-like structures, while tubes tangle and knot 
like spaghetti boiling in a pan (Vincent \& Meneguzzi 1994).  Moreover, vortex tubes usually have short lifetimes, 
vanishing at one place and reforming at another.  The behaviour of Navier-Stokes 
flows diverge in behaviour from Euler flows once viscosity has taken effect in reconnection 
processes. Nevertheless, the creation and early/intermediate evolution of their vortical sets 
appear to be similar. No adequate mathematical theory has been forthcoming explaining why thin 
sets tend to be favoured. 
The purpose of this paper is to investigate this enduring question in the light of the recent 
advances made in in the geometry of K\"{a}hler and other complex manifolds. While many difficult 
questions remain to be solved and explored, we believe that sufficient evidence exists that 
suggests that three-dimensional turbulent vortical dynamics may be governed by geometric principles.

The incompressible Navier-Stokes equations, in two or three dimensions, are
\beq
\label{eq:1} \frac{\pa \bu}{\pa t} + \bu\cdot\nabla\bu +
\frac{1}{\rho}\nabla P = \nu \nabla^{2}\bu\,, 
\eeq 
\beq\label{eq:2}
\frac{\pa \rho}{\pa t} + \nabla\cdot(\rho\bu) = 0\,. 
\eeq 
Here,
$\bu(\bx,t)$ is the fluid velocity, the pressure and density of the fluid are denoted by $P(\xx,t)$
and $\rho(\xx,t)$ respectively, $\nabla$ is the gradient operator and 
$\nu$ is the viscosity; in the inviscid case when $\nu =0$ we have the Euler equations.
The constraint imposed by the incompressibility condition 
\beq
\label{eq:3} 
\nabla\cdot\uu = 0, 
\eeq 
is very severe. It means that the convective derivative of the density vanishes. In turn this means
that an initially homogeneous (constant density) fluid remains constant for all time; 
$\rho(\xx, 0) = \rho(\xx, t) =\mbox{constant}$. Hereafter this density is taken as unity.
Moreover, when (\ref{eq:3}) is applied across (\ref{eq:1}) it demands that velocity derivatives 
and the pressure are related by a Poisson equation
\beq
\label{eq:4}
-\nabla^2 P = u_{i,j}u_{j,i}\,,
\eeq
where $\nabla^2$ is the Laplace operator. Solving (\ref{eq:4}) in tandem
with (\ref{eq:1}) is computationally demanding and is one of the major limiting factors in how far 
numerical calculations can be driven. 

A considerable literature exists on the dynamics of vortex tubes, particularly on the topic of the
Burgers' vortex (Burgers 1948). In an influential paper that contains substantial references, Moffatt,
{\em et al.} (1994) coined the simile \textit{Burgers' vortices are the sinews of turbulence}
and thus identified the heart of the problem; that is, these filament-like vortices stitch together
the large-scale anatomy of vortical dynamics. Despite the twisting, bending and tangling they 
undergo, they appear to be the preferred states of Navier-Stokes turbulent flows. In fact Burgers' 
vortices are examples of a \textit{two-and-a-half-dimensional flow}, which can be defined by the 
class of velocity fields written as (Gibbon {\em et al.} (1999))
\bel{class1}
\bu(x,y,z,t) = \left\{u_{1}(x,y,t),\,u_{2}(x,y,t),\, z\gamma(x,y,t)\right\}\,.
\ee
This flow is linear in $z$ in the $\bk$-direction; thus it is stretching (or compressing) in that 
direction but is linked dynamically to its cross-sectional part. The nomenclature refers to the 
fact that it is neither fully two- nor three-dimensional but lies somewhere 
in-between\footnote{In the case of the three-dimensional Euler equations data can become rough 
very quickly; our
manipulations in this paper are therefore purely formal. In fact it has been shown numerically in 
Ohkitani \& Gibbon (2000) and analytically in Constantin (2000) that solutions of the type in (\ref{class1}) 
can become singular in a finite time, which is consistent with observations that vortex tubes have 
finite life-times; the singularity is not real in the full three-dimensional Euler sense as it has 
infinite energy
but indicates that the flow will not sustain the structure (\ref{class1}) for more than a finite time.
For the possibility of a real Euler singularity see Kerr (1983) and Kerr (2005).} Its components must also 
satisfy the divergence-free condition
\bel{class2}
u_{1,x}(x,y,t)+u_{2,y}(x,y,t) +\gamma(x,y,t) = 0\,.
\ee 
The class of velocity fields in equation (\ref{class1}), first used in Ohkitani \& Gibbon (2000), is a more general
classification of Burgers-type solutions and contains the specific form of the Burgers vortex solutions
used in Moffat {\em et al.} (1994).
Included in (\ref{class1}) are the Euler solutions of Stuart (1987, 1991),
in which $u_{1}$ and $u_{2}$ are also linear in $x$ (say) leaving the dependent variables to be functions
of $y$ and $t$. Then stretching can occur in two directions thereby producing sheet-like vortical solutions.

The differences between the three-dimensional and two-dimensional Navier-Stokes equations are 
fundamental because the vortex stretching term $\bw\cdot\nabla\bu$ in the equation for vorticity
is present in the former but absent in the latter. Nevertheless, Lundgren (1982) has shown that for \twohd
flows of the type 
\bel{Lun1}
u_{1} = -\shalf x\gamma(t)+\psi_{y}~~~~~~u_{2} = -\shalf y\gamma(t)-\psi_{x}~~~~~u_{3}=z\gamma(t)
\ee
can be mapped into solutions of the two-dimensional Navier-Stokes equations with $\psi(x,y,t)$ as 
a stream function.  

To investigate the geometric structure behind these solutions requires certain technical tools; 
these are outlined in \S2 of this paper. The constraint in equation (\ref{eq:4}) is the basis of our geometric arguments, and because it is true 
for both the Navier-Stokes and Euler equations,
the conclusions reached in this paper are valid for both cases. It is, of course, to be expected 
that any geometric structure should be independent of viscosity. From now on when we refer to the 
Navier-Stokes equations it should be implicitly understood that the Euler equations are also 
included. The K\"ahler structure for the two-dimensional Navier-Stokes
equations is described in \S3 and then formulated for \twohd Navier-Stokes flows in \S4. 
Our results show that the necessary condition on the pressure for a K\"{a}hler structure 
to exist in two spatial dimensions (with time entering only as a parameter) for the two-dimensional
Navier-Stokes equations is $\nabla^{2}P> 0$. This constraint is highly restrictive: by no means all
two-dimensional Navier-Stokes flows would conform to it. More promising is the equivalent condition
for two-and-a-half dimensional solutions of type (\ref{Lun1}). Theorem \ref{3dthm} in \S\ref{S4} 
shows that these \twohd solutions have an underlying K\"{a}hler structure if $\nabla^{2}P$ has a 
very large negative lower bound, thus associating a wide set of `thin' solutions with the K\"{a}hler
property. While the existence of a negative finite lower bound suggests some work still needs to 
be done, this result implies that preferred vortical thin sets have a connection with 
a K\"{a}hler geometric structure that deserves further study.

The solutions considered in this paper represent the ideal cases of straight tubes or flat 
sheets; in reality, as indicated in the first paragraph of this section, these vortical 
objects constantly undergo processes of bending and tangling. Speculatively, it is possible 
that once this process is underway, solutions move from living on a  K\"{a}hler manifold in 
two complex dimensions to other complex manifolds of a higher dimension, although this is a 
much more difficult mathematical problem to address. In \S\ref{S5} and Appendix \ref{A1}, 
following Banos (2002), we describe an explicit connection between Monge--Amp\'{e}re operators 
and complex manifolds of Calabi--Yau type which are associated with a restricted class of 
three-dimensional Navier-Stokes flows. This explicit connection exists only for a very special class of 
flows that require the pressure to satisfy 
$\nabla^{2}P < 0$. What this limited class of flows physically represent is unclear at 
the present time.

The work of Roubtsov \& Roulstone (1997, 2001) showed how quaternionic and hyper-K\"{a}hler 
structures emerge in models of nearly geostrophic flows in atmosphere and ocean 
dynamics. These results were based on earlier work by McIntyre and Roulstone and was reviewed 
by them in McIntyre \& Roulstone (2002).  It has also been shown that the three-dimensional Euler equations 
has a quaternionic structure in the dependent variables (Gibbon 2002). The use of different sets of dependent and independent variables in geophysical models of cyclones and fronts, has facilitated some remarkable simplifications of otherwise hopelessly difficult nonlinear problems (see, for example, Hoskin \& Bretherton 1972). Roulstone \& Sewell (1997) and McIntyre \& Roulstone (2002), describe how contact and K\"ahler geometries provide a framework for understanding the basis of the various coordinate transformations that have proven so useful in this context. This present work has evolved from revisiting Gibbon (2002) in the light of Roubtsov \& Roulstone (2001).

\section{Differential Forms and Monge--Amp\`ere Equations}\label{S2}

In this section we prepare some tools that enable us to study certain partial differential equations
arising in incompressible Navier-Stokes flows from the point-of-view of differential geometry. An 
introduction to the application of some basic elements of exterior calculus to the study of partial
differential equations, with application to fluid dynamics, can be found in McIntyre \& Roulstone (2002). Here, we
shall draw largely on Lychagin {\em et al.} (1993) and Banos (2002).

A Monge--Amp\`ere equation (MAE) is a second order partial differential equation, which, 
for instance in two variables, can be written as follows:
\begin{equation}{\label{eq:MA2}}
A\phi_{xx}+2B\phi_{xy}+C\phi_{yy}+D(\phi_{xx}\phi_{yy}-\phi_{xy}^2)+E=0,
\end{equation}
where $A,B,C$ and $D$ are smooth functions of $(x,y,\phi, \phi_x,\phi_y)$. This equation is 
elliptic if
\beq
\label{eq:ec}
AC - 4B^2 - DE > 0.
\eeq
In dimension $n$, a
Monge--Amp\`ere equation is a linear combination of the minors of the hessian matrix\footnote{We 
denote by $\hess(\phi)$ the determinant of the hessian matrix of $\phi$. For example, in two
variables, $\hess(\phi)=\phi_{xx}\phi_{yy}-\phi_{xy}^2$.} of $\phi$. We shall refer to such 
equations as {\em symplectic} MAEs when the coefficients $A,B,C$ and $D$ are smooth functions 
of $(x,y, \phi_x,\phi_y)\in T^*\mathbb{R}^2$; i.e. they are smooth functions on the quotient 
bundle $J^1\mathbb{R}^2/J^0\mathbb{R}^2$, where $J^1\mathbb{R}^2$ denotes the manifold of 
1-jets on $\mathbb{R}^2$.

\subsection{Monge--Amp\`ere operators}

Lychagin (1979) has proposed a geometric approach to these equations, using differential forms 
on the cotangent space (i.e. the phase space). The idea is to associate with a 
form\footnote{The use of the Greek letters $\omega$ and $\Omega$ is common in differential 
geometry; these symbols should not be confused with the fluid vorticity vector $\bw$.}
$\omega\in
\Lambda^n(T^*\mathbb{R}^n)$, where $\Lambda^n$ denotes the space of differential $n$-forms 
on $T^*\mathbb{R}^n$, the Monge--Amp\`ere equation $\triangle_\omega=0$, where  
$\triangle_{\omega} : C^\infty(\mathbb{R}^n)\rightarrow
\Omega^n(\mathbb{R}^n)\cong C^\infty(\mathbb{R}^n)$ is the
differential operator, $\triangle_\omega$, defined by
$$
\triangle_{\omega}(\phi)=(\drm\phi)^*\omega\,.
$$
We denote by $(\drm\phi)^*\omega$ the restriction of $\omega$ to the graph of $\drm\phi$ 
($\drm\phi: \mathbb{R}^n\rightarrow T^*\mathbb{R}^n$ is the differential of $\phi$).  
A form $\omega\in \Lambda^n(T^*\mathbb{R}^n)$ is said to be
effective if $\omega \wedge\Omega=0$, where $\Omega$ is the canonical symplectic form on 
$T^*\mathbb{R}^n$. Then the 
so called Hodge-Lepage-Lychagin theorem tells us that this correspondence between MAEs and 
effective forms is one to one. For instance, the Monge--Amp\`ere equation $\eqref{eq:MA2}$ is associated 
with the effective form
$$
\omega = A\drm p\wedge \drm y +B(\drm x\wedge \drm p-\drm y\wedge \drm q)+C\drm x\wedge \drm q + 
D\drm p\wedge\drm q +E\drm x\wedge \drm y,
$$
where $(x,y,p,q)$ is the symplectic system of coordinates of
$T^*\mathbb{R}^2$, and on the graph of $\drm\phi,\;\;p=\phi_x\;\mbox{and}\; q=\phi_y$. So, for example, if we pull-back the one-form $\drm p$ to the base space, we have $\drm p = \phi_{xx}\drm x + \phi_{xy}\drm y$, and then $\drm p\wedge \drm q = \hess(\phi)\drm x \wedge \drm y$, where we have also used the skew symmetry of the wedge product.

\subsection{Monge--Amp\`ere structures}

The geometry of MAEs in $n$ variables can be described by a  pair $(\Omega,\omega)\in
\Lambda^2(T^*\mathbb{R}^n)\times \Lambda^n(T^*\mathbb{R}^n)$ such that
\begin{enumerate}
\item $\Omega$ is symplectic; that is, nondegenerate ($\Omega\wedge\Omega\neq 0$) and 
closed ($\drm\Omega = 0$)

\item $\omega$ is effective; that is, $\omega\wedge\Omega=0$.
\end{enumerate}
Such a pair is called a Monge--Amp\`ere structure.
In four dimensions (that is $n=2$), this geometry can be either complex or real and this 
distinction coincides with the usual distinction between elliptic and hyperbolic, respectively, 
for differential equations in two variables. Indeed,  when $\omega\in
\Lambda^2(T^*\mathbb{R}^2)$ is a non-degenerate $2$-form
($\omega\wedge\omega\neq 0$), one can associate with  the
Monge--Amp\`ere structure $(\Omega,\omega)\in
\Lambda^2(T^*\mathbb{R}^2)\times \Lambda^2(T^*\mathbb{R}^2)$ the
tensor $I_\omega$ defined by
$$
\frac{1}{\sqrt{|\pf(\omega)|}}\omega(\cdot,\cdot)=\Omega(I_\omega\cdot,\cdot)
$$
where $\pf(\omega)$ is the pfaffian of $\omega$:
$\omega\wedge\omega = \pf(\omega)(\Omega\wedge\Omega)$. Thus, for the effective form $\omega$ 
associated with the MAE (\ref{eq:MA2}), the $\pf(\omega)$ coincides with (\ref{eq:ec}). This
tensor is either an almost complex structure or an almost product structure:
\begin{enumerate}
\item $\triangle_\omega$ is elliptic $\Leftrightarrow$
$\pf(\omega)>0$ $\Leftrightarrow$ $I_\omega^2=-Id$
\item $\triangle_\omega$ is hyperbolic $\Leftrightarrow$
$\pf(\omega)<0$ $\Leftrightarrow$ $I_\omega^2=Id$
\end{enumerate}
and it is integrable if and only if
\beq
\label{eq:intcond}
\drm\left(\frac{1}{\sqrt{|\pf(\omega)|}}\omega\right) = 0.
\eeq
Given a pair of two-forms $(\Omega,\omega)$ on $T^*\mathbb{R}^2$, such that $\omega\wedge\Omega=0$,
by fixing the volume form in terms of $\Omega$, we can define a pseudo-riemannian metric $g_\omega$
in terms of the quadratic form
\beq
\label{eq:gomega}
g_\omega(X,Y) = \frac{\iota_X\Omega\wedge\iota_Y\omega + \iota_Y\Omega\wedge\iota_X\omega}{\Omega\wedge\Omega}\wedge
\pi^*(vol),\;\;X,Y\in T\mathbb{R}^2 ,
\eeq
where $vol$ is the volume form on $\mathbb{R}^2$ and $\pi:T^*\mathbb{R}^{2}\mapsto\mathbb{R}^{2}$. We
can now identify our MAE given by $\omega$ with an almost K\"ahler structure given by the triple 
$(\mathbb{R}^{2}, g_{\omega}, I_{\omega})$ via 
\beq
\omega(X,Y) \equiv g_{\omega}(I_{\omega} X,Y) .
\eeq
One can go further and in $\mathbb{R}^4$, one can show how a natural hyper-K\"ahler 
structure emerges by identifying points in $\mathbb{R}^4$ with quaternions $\ell\in\mathbb{H}$. This
structure was utilised by Roubtsov \& Roulstone (1997, 2001) in their description of nearly geostrophic
models of meteorological flows.

\section{A K\"ahler Structure for two-dimensional Navier-Stokes flows}\label{S3}

We shall now show Monge--Amp\`ere equations, in two independent variables (sat $(x,y)\in\mathbb{R}^2$),
arise in two-dimensional and \twohd incompressible Navier-Stokes flows. This in turn, via the relationships
described in the previous section, leads us to a K\"ahler structure for these flows.

If the flow described by ({\ref{eq:1}) is two-dimensional, and the fluid is incompressible,
we can represent the velocity by \beq \uu = \kk\times\nabla\psi,
\eeq where $\psi(x,y,t)$ is a stream function and $\kk$ is the
local unit vector in the vertical. If we substitute this for the
velocity in (\ref{eq:4}), we get 
\beq \label{eq:5} 
\nabla^2 P =
-2(\psi^2_{xy} - \psi_{xx}\psi_{yy}) \eeq 
 This is an equation of Monge--Amp\`ere type (cf. (\ref{eq:MA2})) for
$\psi$, given $
\nabla^2 P (=P_{xx}+P_{yy})$, and it is an elliptic Monge--Amp\`ere equation (cf. (\ref{eq:MA2}) and (\ref{eq:ec}) with
$E=\nabla^2 P, D = -2, A=B=C=0$) if
\beq
\label{eq:irp}
\nabla^2 P > 0 .
\eeq
Following, for example, Lychagin \& Roubtsov (1983), we introduce
the usual notation for coordinates on $T^*\mathbb{R}^2$, $p = \psi_x,
q = \psi_y$, and then we can write (\ref{eq:5}) on the graph of
 $\drm\psi$ via
\beq
\label{eq:5a} \omega_{2d}\equiv
\nabla^2 P\:\drm x\wedge\drm y - 2\drm p\wedge\drm q;\;\;\triangle_{\omega_{2d}}=0 .
\eeq 
We have also on the graph of $\drm\psi$
\begin{equation}
\label{Cauchy}
 \Omega\equiv\drm x\wedge \drm p + \drm y \wedge \drm q;\;\; \triangle_\Omega =0,
\end{equation}
which says simply that $\psi_{xy}=\psi_{yx}$. Equations (\ref{eq:5a}) and (\ref{Cauchy})
define an almost complex structure, $I_{\omega_{2d}}$, on $T^*\mathbb{R}^2$, given in coordinates by
$$
I_{ik}=\frac{1}{\sqrt{2
\nabla^2 P}} \Omega_{ij}\omega_{jk} .$$ 
 We have
$$
I_{ij}=\begin{pmatrix} 0&0&0&-\frac{1}{\alpha}\\
0&0&\frac{1}{\alpha}&0\\
0&-\alpha&0&0\\
\alpha&0&0&0\\
\end{pmatrix}
$$
with $
\nabla^2 P= 2\alpha^2$. This almost complex structure is integrable (cf. (\ref{eq:intcond})) in the special case
\beq
\label{eq:pconst}
\nabla^2 P = constant\,.
\eeq
Recall that time is merely a parameter here. When $P$ satisfies (\ref{eq:pconst}), we can introduce the coordinates
${\cal X,Y}$, and a two-form $\omega_{{\cal XY}}$
\beq {\cal
X} = x - i\alpha^{-1}q,\;\;{\cal Y} = y + i\alpha^{-1}p,\;\;\omega_{{\cal XY}} = \drm {\cal X}\wedge
\drm {\cal Y},
\label{eq:6} 
\eeq  
then (\ref{eq:5}) together with (\ref{Cauchy}) are equivalent to 
\beq 
\triangle_{\omega_{{\cal XY}}} = 0.
\eeq 
To summarize, the graph of $\psi$ is a complex curve in $(T^*\mathbb{R}^2,I_{\omega_{2d}})$. This is
a starting point for a K\"ahler description of the incompressible two-dimensional Navier-Stokes 
equations.

\section{A result for two-and-a-half dimensional flows}\label{S4}

At this point it is appropriate to work with the \twohd Burgers solutions 
introduced in \S\ref{S1} in equations (\ref{class1}), (\ref{class2}) and (\ref{Lun1}). 
Based on the results of the last section, we shall prove a more realistic result 
for \twohd flows in Theorem \ref{3dthm}. 

Lundgren (1982) made a significant advance when he showed that the class of three-dimensional
Navier-Stokes solutions (designated in \S1 as two-and-a-half-dimensional) 
\bel{twohalf}
u_{1}(x,y,t) = -\shalf\gamma(t)x + \psi_{y}\,;\hspace{2cm}
u_{2}(x,y,t) = -\shalf\gamma(t)y - \psi_{x}
\ee
\bel{twohalfa}
u_{3}(x,y,t) = z\gamma(t) + \phi(x,y,t)
\ee
under the limited conditions of a constant strain $\gamma(t)=\gamma_{0}$, can be mapped back to 
the two-dimensional Navier-Stokes equations under a stretched co-ordinate transformation; see also 
Majda (1986), Majda \& Bertozzi (2002), Saffman (1993), and Pullin \& Saffman (1998).  In (\ref{twohalf}), $\psi=\psi(x,y,t)$ is a two-dimensional stream 
function. This idea was extended by Gibbon \textit{et al.} (1999) to a time-dependent strain 
field $\gamma = \gamma(t)$ with the inclusion of a scalar $\phi(x,y,t)$ in (\ref{twohalfa}). The 
class of solutions in (\ref{twohalf}), which are said to be of \textit{Burgers-type}, is generally 
thought to represent the observed tube-sheet class of solutions in Navier-Stokes turbulent 
flows (Moffat {\em et al.} (1994) and Vincent \& Meneguzzi (1994)). 

Depending upon the sign of $\gamma(t)$ the vortex represented by (\ref{twohalf}) either stretches in
the $z$-direction and contracts in the horizontal plane, which is the classic Burgers vortex tube, 
or vice-versa, which produces a Burgers' vortex shear layer or sheet. Thus $\gamma$, which can be 
interpreted as the aggregate effect of other vortices in the flow, acts an externally imposed strain
function or `puppet master', and can switch a vortex between the two extremes of these two 
topologies as we discussed in \S1.

This class of solutions is connected to the results of \S2 through the following theorem, 
which is the main result of this section, and of the paper\footnote{The notation used in this section is: 
$\nabla$ is the two-dimensional gradient and $\nabla_{3}$ is the three-dimensional gradient. 
$\nabla^{2}$ and $\nabla^{2}_{3}$ are the two- and three-dimensional Laplacians respectively 
(to avoid confusion with the symbol $\triangle$ in \S2).}:
\begin{theorem}\label{3dthm}
If a two-and-a-half-dimensional Burgers-type class of solutions has a Laplacian of the pressure 
that is bounded by $\nabla^{2}_{3}P > -\thhalf\gamma^{2}$ then any associated underlying 
two-dimensional Navier-Stokes flow is of K\"{a}hler type.
\end{theorem}
\par\medskip\noindent
\textbf{Proof:} To prove this theorem we first need two Lemmas. Firstly let 
$\bu = \left(u_{1},u_{2},u_{3}\right)$ 
be a candidate velocity field solution of the three-dimensional Navier-Stokes equations taken in 
the form
\bel{AA1}
u_{1} = u_{1}(x,y,t)
\hspace{1cm}
u_{2} = u_{2}(x,y,t)
\hspace{1cm}
u_{3} = z\gamma(x,y,t) + \phi(x,y,t).
\ee
with $z$ appearing only in $u_{3}$.  With this velocity field the total derivative is now
\bel{AA4}
\frac{\Drm~}{\Drm t} = \frac{\partial~}{\partial t} 
+ u_{1}\frac{\partial~}{\partial x}
+ u_{2}\frac{\partial~}{\partial y}
+ (z\gamma + \phi)\frac{\partial~}{\partial z}
\ee
and the vorticity vector $\bw$ must satisfy
\bel{ns3}
\frac{\cD\bw}{\cD t} = S\bw + \nu\nabla^{2}\omega\,,
\ee
where $S$ is the strain matrix whose elements are $S_{ij}= \shalf\left(u_{i,j}+ u_{j,i}\right)$.
In the following Lemma $\bv(x,y,t) = (u_{1},u_{2})$, and $\mathcal{P}(x,y,t)$ is a two-dimensional
pressure variable which is related to the full pressure $P$ in (\ref{A8}). The material derivative 
is now
\bel{AA5}
\frac{\cD~}{\cD t} = \frac{\partial~}{\partial t} +\bv\cdot\nabla
\ee
\begin{lemma} (see Gibbon {\em et al.} 1999) Consider the velocity field $\bu =(\bv,\,z\gamma +\phi)$; then 
$\bv$, $\omega_{3}$, $\phi$ and $\gamma$ satisfy
\bel{prop1a}
\frac{\cD\bv}{\cD t} + \nabla\mathcal{P} = \nu\nabla^{2}\bv
\hspace{2cm}
\frac{\cD\omega_{3}}{\cD t} = \gamma \omega_{3} + \nu\nabla^{2}\omega_{3}\,,
\ee
\bel{prop2}
\frac{\cD\phi}{\cD t} = -\gamma \phi + \nu\nabla^{2}\phi\,,
\ee
\bel{prop3}
\frac{\cD\gamma}{\cD t} + \gamma^{2} + P_{zz}(t) = \nu\nabla^{2}\gamma\,.
\ee
The velocity field $\bv$ satisfies the continuity condition $\div\,\bv = - \gamma$ and the 
second partial $z$-derivative of the pressure $P_{zz}$ is constrained to be spatially uniform.
\end{lemma}
\par\medskip\noindent
\textbf{Remark:} While (\ref{prop1a}) 
looks like a two-dimensional Navier-Stokes flow, the continuity condition implies that the 
two-dimensional divergence $\mbox{div}\bv \neq 0$; thus an element of three-dimensionality 
remains.  
\par\medskip\noindent
\textbf{Proof:}  The evolution of the third velocity component $u_{3} = \gamma z + \phi$ is 
given by
\bel{A6}
- P_{z}= \frac{\Drm u_{3}}{\Drm t} - \nu\nabla^{2}u_{3} = 
z \left(\frac{\cD\gamma}{\cD t} + \gamma^{2}-\nu\nabla^{2}\gamma\right) 
+ \left(\frac{\cD\phi}{\cD t} + \gamma\phi -\nu\nabla^{2}\phi\right)
\ee
which, on integration with respect to $z$, gives 
\bel{A8}
- P(x,y,z,t) = \shalf z^{2}\left(\frac{\cD\gamma}{\cD t}+\gamma^{2}-\nu\nabla^{2}\gamma\right)  
+ z\left(\frac{\cD\phi}{\cD t} + \gamma\phi -\nu\nabla^{2}\phi\right) - \mathcal{P} (x,y,t).
\ee
It is in this way that $\mathcal{P} (x,y,t)$ is related to $P(x,y,z,t)$. 
However, from the first two components of the Navier-Stokes equations, we know that $\nabla P$ 
must be independent of $z$. For this to be true the coefficients of $z$  
and $z^2$ in (\ref{A8}) must necessarily satisfy
\bel{A9}
\frac{\cD\phi}{\cD t} + \gamma\phi  - \nu\nabla^{2}\phi= c_{1}(t)
\hspace{2cm}
\frac{\cD\gamma}{\cD t} + \gamma^{2} -\nu\nabla^{2}\gamma = c_{2}(t).
\ee
$c_{1}(t)$ is an acceleration of the co-ordinate frame which can be taken as zero without 
loss of generality. Equation (\ref{A8}) shows  that  $c_{2}(t) = - P_{zz}(t)$ which restricts  
$P_{zz}$ to being spatially uniform. To find the evolution of $\omega_{3}$ we consider the 
strain matrix $S =\{S_{ij}\}$ 
\bel{A11}
S = \left(
\begin{array}{ccc}
u_{1,x} & \frac{1}{2}\left(u_{1,y}+u_{2,x}\right)
&\shalf\left( z\gamma_{x} + \phi_{x}\right)\\
\shalf\left(u_{1,y}+u_{2,x}\right)& u_{2,y}&
\shalf\left( z\gamma_{y} + \phi_{y}\right)\\
\shalf\left(z\gamma_{x} + \phi_{x}\right)&
\shalf\left(z\gamma_{y} + \phi_{y}\right)& \gamma
\end{array}
\right).
\ee
Working out the vorticity field $\bw$ from (\ref{AA1}) it is easily seen that $(S\bw)_{3} = \gamma\omega_{3}$.
Thus (\ref{ns3}) shows that $\omega_{3}$ decouples from $\phi$ to give the equation for 
$\omega_{3}$ in (\ref{prop1a}). \hspace{1cm}$\square$
\par\bigskip
Now let us consider the class of Burgers' velocity fields given in 
(\ref{twohalf}) with a stream function $\psi(x,y,t)$. The strain rate variable 
$\gamma$ is taken as a function of time only.  The continuity condition 
is now automatically satisfied.  The material derivative is given by
\bel{total3}
\frac{\cD~}{\cD t} = \frac{\partial~}{\partial t} - \shalf\gamma(t)\left(x
\frac{\partial~}{\partial x}  + y \frac{\partial~}{\partial y}\right) 
+ J_{x,y}(\psi,\,\cdot).
\ee
New co-ordinates can be taken (Lundgren's transformation (Lundgren (1982))) 
\bel{NS6}
s(t) = \exp\left(\int_{0}^{t}\gamma(t')\,dt'\right)
\ee
\bel{NS7}
\tilde{x} = s^{1/2}x
\hspace{1.5cm}
\tilde{y} = s^{1/2}y
\hspace{1.5cm}
\tilde{t} = \int_{0}^{t}s(t')\,dt',
\ee
which re-scale $\omega_{3}$ and $\phi$ into new variables
\bel{nsextra}
\tilde{\omega}_{3}(\tilde{x},\tilde{y},\tilde{t}) = s^{-1}\omega_{3}(x,y,t)\hspace{1cm}
\hspace{2cm}
\tilde{\phi}(\tilde{x},\tilde{y},\tilde{t}) = s\,\phi(x,y,t).
\ee
The material derivative is
\bel{NS10}
\frac{\cD~}{\cD\tilde{t}} = \frac{\partial~}{\partial\tilde{t}} + \bvt\cdot\tilde{\nabla}
\ee
where $\psi(x,y,t) = \tilde{\psi}(\tilde{x},\tilde{y},\tilde{t})$; 
$\bvt = \left(\tilde{\psi}_{\tilde{y}}\,,-\tilde{\psi}_{\tilde{x}}\right)$ 
and 
$\tilde{\nabla} = \bi\,\partial_{\tilde{x}} +\bj\,\partial_{\tilde{y}}$.  The relation 
between $\bv =(u_{1},\,u_{2})$ and $\bvt$ is given by
\bel{vdef}
u_{1} = -\shalf\gamma(t) x + s^{1/2}\tilde{v}_{1}\,;
\hspace{2cm}
u_{2} = -\shalf\gamma(t) y + s^{1/2}\tilde{v}_{2}
\ee
and the relation between the two material derivatives in combination with the respective 
Laplacians is
\bel{mat1}
\frac{\cD~}{\cD t} -\nu\nabla^{2} = s\left(\frac{\cD~}{\cD\tilde{t}}-\nu\tilde{\nabla}^{2}\right)\,.
\ee
Introducing a new pressure variable $\tilde{P}$ as 
\bel{newP}
\tilde{P} = s^{-1}
\left[\mathcal{P} - \squart(x^{2}+y^{2})\left(\dot{\gamma} - \shalf\gamma^{2}\right)\right]
\ee 
our results can be summarized in our second Lemma:
\begin{lemma}\label{GFDthm}
The re-scaled velocity field $\bvt$ satisfies the two-dimensional re-scaled Navier-Stokes 
equations ($\div\,\bvt = 0$)
\bel{NS9a}
\frac{\cD\bvt}{\cD\tilde{t}} + \tilde{\nabla}\tilde{P}= \nu\tilde{\nabla}^{2}\bvt\,.
\ee
The vorticity $\tilde{\omega}_{3}(\tilde{x},\tilde{y},\tilde{t}) 
= -\tilde{\nabla}^{2}\,\tilde{\psi}$ and the passive scalar 
$\tilde{\phi}(\tilde{x},\tilde{y},\tilde{t})$ satisfy
\bel{NS9b}
\frac{\cD\tilde{\omega}_{3}}{\cD\tilde{t}} = \nu\tilde{\nabla}^{2}\omega_{3}\,,
\hspace{2cm}
\frac{\cD\tilde{\phi}}{\cD\tilde{t}} = \nu\tilde{\nabla}^{2}\phi\,.
\ee
\end{lemma}
\par\medskip\noindent
\textbf{Proof:}  From (\ref{NS6}) we note the useful result that $\Drm s/\Drm t = \gamma s$. 
Using (\ref{vdef}) we write
\bel{lem2a}
\frac{\Drm u_{1}}{\Drm t} - \nu\nabla^{2}u_{1} 
= -\shalf x\left(\dot{\gamma} - \shalf\gamma^{2}\right) + 
s^{3/2}\left(\frac{\cD \tilde{v}_{1}}{\cD\tilde{t}} - \nu\nabla^{2}\tilde{v}_{1}\right)\,,
\ee
\bel{lem2b}
\frac{\Drm u_{2}}{\Drm t} - \nu\nabla^{2}u_{2} 
= -\shalf y\left(\dot{\gamma} - \shalf\gamma^{2}\right) -
s^{3/2}\left(\frac{\cD \tilde{v}_{2}}{\cD\tilde{t}} - \nu\nabla^{2}\tilde{v}_{2}\right)\,.
\ee
Next we appeal to the definition of the pressure $\tilde{P}$ in (\ref{newP}) to give the velocity 
pressure relation in (\ref{NS9a}). The results for $\tilde{\phi}$ and $\tilde{\omega}_{3}$ 
follow immediately. \hspace{2cm}$\square$
\par\medskip
The proof of Theorem \ref{3dthm} is now ready to be completed. To obtain the full three-dimensional
Laplacian of the pressure $\nabla^{2}_{3} P$ we use (\ref{newP}) and (\ref{prop3}) and write 
\beqa{Lap1}
- \nabla^{2}_{3}P &=& \thhalf\gamma^{2} +
s^{2}\left[\frac{\partial~}{\partial\tilde{x}}\left(\frac{\cD \tilde{v}_{1}}{\cD\tilde{t}} - \nu\nabla^{2}\tilde{v}_{1}\right)
+ \frac{\partial~}{\partial\tilde{y}}\left(\frac{\cD \tilde{v}_{2}}{\cD\tilde{t}} - \nu\nabla^{2}\tilde{v}_{2}\right)
\right]\nonumber\\
&=& \thhalf\gamma^{2} - s^{2}\tilde{\nabla}^{2}\tilde{P}\,.
\eq
Thus if $\nabla^{2}_{3}P$ satisfies the condition in Theorem \ref{3dthm} then the 
corresponding K\"{a}hler
positivity condition (\ref{eq:irp}) on the Laplacian for two-dimensional flow is satisfied.
\hspace{1cm}$\blacksquare$
\par\medskip
Lundgren's mapping breaks down under one condition: while the strain $\gamma(t)$ can take either sign,
if it is forever negative or for long intervals, the domain $t\in[0,\,\infty]$ maps on to a finite 
section of the $\tilde{t}$-axis.
For example, if $\gamma = -\gamma_{0} =\mbox{const}<0$
then $s = \exp(-\gamma_{0}t)$  and $\tilde{t} = \gamma_{0}^{-1}\left[1- \exp(-\gamma_{0}t)\right]$.
Hence $t \in[0,\,\infty]$  maps onto $\tilde{t} \in[0,\,\gamma_{0}^{-1}]$.

\section{Remarks on more general 3D Navier-Stokes flows}\label{S5}

One of the features of this paper has been the discussion between the correspondence between 
Burgers-like flows and K\"{a}hler geometry: given the ubiquity of such flows it is reasonable 
to think of them as a `preferred' or `attracting' class of solutions.  In \S\ref{S1} we speculated 
on the possibility that once the flow moves away from this class of preferred solutions then they 
may move onto complex manifolds of a higher dimension. One of the simplest of these is the 
Calabi--Yau manifold. These are smooth, compact, complex manifolds with a Ricci-flat K\"{a}hler 
metric with a holomorphically trivial canonical bundle. 

To investigate the question of other manifolds exist in more general three-dimensional Navier-Stokes
flows we exploit the fact that the incompressibility condition allows us to write $\bu$ in
terms of a vector potential $\AA$
\beq
\label{vp}
\uu = \curl\AA\,.
\eeq
We might suppose that the vector potential is some function of a scalar potential $\psi(\xx,t)$
\beq\label{vecpot1}
\AA = (F(\xx, t,\psi(\xx,t)), G(\xx,t,\psi(\xx,t)), H(\xx,t,\psi(\xx,t)))\,,
\eeq
where $F$, $G$ and $H$ are any (appropriately differentiable) functions and our notation indicates that
they may have an explicit dependence on $\xx$ and $t$. If we insert such a flow, using (\ref{vp}), into
(\ref{eq:4}), then we obtain, in general, a variable-coefficient Monge--Amp\`ere equation for $\psi$.
As we explain in the Appendix (following Banos 2002), there is a correspondence between Monge--Amp\`ere
equations and Calabi--Yau geometries in six dimensions.  Calabi--Yau geometries have been much studied
in connection with string theory and in the Appendix we furnish an example of how such a geometry 
arises in the three-dimensional Navier-Stokes equations, in the case of a very special choice
of vector potential.

A further variation on this theme revolves around the addition of rotation to the system,
which has important meteorological applications. The Euler equations of motion now become 
\beq
\label{eq:48a}
\frac{\pa \vv}{\pa t} + \vv\cdot\nabla\vv + f(\kk\times\vv) + \nabla P = 0\,,
\eeq 
where $\shalf f$ is the angular frequency. If we examine these equations in two dimensions, with constant rotation, then 
taking the divergence of (\ref{eq:48a}) gives 
\beq\label{eq:7} 
\nabla^2 P = - 2(\psi^2_{xy} -
\psi_{xx}\psi_{yy}) + f\nabla^2\psi . 
\eeq 
This is an elliptic Monge--Amp\`ere equation if $\nabla^2 P + f^2/2 >0$\,. The associated complex structure is integrable when $\nabla^2 P$  is a constant (cf. (\ref{eq:intcond})), and in this case
we can introduce new complex coordinates
\beq
\label{eq:XYf}
\tilde{X}  =  a x +i(fy+2q),\;\;
\tilde{Y}  =  a y -i(fx+2p)
 \eeq
 with $a=(2\nabla^2 P +f^2)^{1/2}$. Once again, (\ref{eq:7}) together with
(\ref{Cauchy}) are equivalent to
$$
\omega_{\tilde{X}\tilde{Y}}\equiv\drm \tilde{X}\wedge 
\drm \tilde{Y},\;\;\triangle_{\omega_{\tilde{X}\tilde{Y}}}=0.
$$
If the pressure is zero, or harmonic, then (\ref{eq:7}) is suggestive of a special Lagrangian
structure. A special Lagrangian structure has also been noted in the work of Roubtsov \& Roulstone (2001), but its role 
in that context is obscure (see McIntyre \& Roulstone (2002) equation (13.27) {\em et seq.}).

\section{Summary}

We have shown how K\"ahler geometry arises in the Navier-Stokes equations of incompressible hydrodynamics, via a 
Monge--Amp\`ere equation associated with (\ref{eq:4}). Although it is certainly not the case that all two-dimensional flows will satisfy the condition for the K\"ahler structure to exist, the situation looks much more promising for two-and-a-half-dimensional flows, of which Burgers vortex is one example. In more general three-dimensional flows, there is a link with geometries of Calabi--Yau type, but this connection requires further investigation.

Issues relating to the existence and interpretation of K\"ahler structures, the integrability conditions, and related matters involving contact and symplectic structures, were discussed by McIntyre \& Roulstone (2002) in connection with various Monge--Amp\`ere equations arising in geophysical fluid dynamics. The semi-geostrophic equations of meteorology, which are a particularly useful model for studying the formation of fronts, were the starting point in McIntyre \& Roulstone {\em op. cit.} for an investigation into the role of novel coordinate systems, similar to those we have found here in (\ref{eq:6}) and (\ref{eq:XYf}). In semi-geostrophic theories, such coordinates facilitate significant simplifications of difficult nonlinear problems, and they are associated with canonical Hamiltonian formulations of these systems. Issues relating to contact and symplectic geometry may also be relevant to the results presented in this paper, and this suggests one direction for further study.

Finally, it is perhaps appropriate to make a comment here about compressible flows. In two dimensions, a velocity potential $\chi(\xx,t)$ may be introduced, so that in (\ref{eq:1}) and (\ref{eq:2}) we have $\uu(\xx,t) = \nabla\chi + \kk\times\nabla\psi$. We may take the divergence of (\ref{eq:1}) and this yields an equation for $\psi$ of the form
\beq
\label{eq:final}
2(\psi_{xy}^2 - \psi_{xx}\psi_{yy}) + 2\chi_{xy}(\psi_{xx} - \psi_{yy}) = R,
\eeq
where $R$ is a function of $\rho$ and $P$, and their spatial derivatives; it is also a function of the first spatial derivatives of $\psi$, and it will also depend on the derivatives of $\chi$ with respect to space and time (e.g. $\chi_x, \chi_{xy}, \chi_{xxt}$ and $\chi_{yyy}$ etc.). In other words, we have a variable-coefficient Monge--Amp\`ere-type equation for $\psi$, when $\rho, P$ and $\chi$ (and their derivatives) are known. It is therefore possible, in principle, that a 
K\"ahler structure may be associated with this type of equation; however, in practice, such a structure will be difficult to identify.


\appendix
\section{Calabi--Yau structures and $3D$-Navier-Stokes flows}\label{A1}

In six dimensions ($T^*\mathbb{R}^3$), there is again a correspondence between
real/complex geometry and nondegenerate Monge--Amp\`ere
structures (Banos 2002). A Monge--Amp\`ere structure $(\Omega,\omega)\in
\Lambda^2(N^6)\times \Lambda^3(N^6)$ on a $6$-dimensional manifold
$N$ ($N=T^*\mathbb{R}^3$) is said to be nondegenerate if $\omega$
is nondegenerate in the sense of Hitchin (2000): the
Hitchin pfaffian $\lambda(\omega)\in C^\infty(N)$ never vanishes 
(cf. (\ref{eq:cy1})-(\ref{eq:cy3}) below).
We note that:
\begin{enumerate}
\item $\lambda(\omega)>0$ if and only if $\omega$ is the sum of two decomposable forms, that is
$$
\omega=\alpha_1\wedge\alpha_2\wedge\alpha_3+\beta_1\wedge\beta_2\wedge\beta_3
$$
\item $\lambda(\omega)<0$ if and only if $\omega$ is the sum of two decomposable complex forms,
that is
$$
\omega=
(\alpha_1+i\beta_1)\wedge(\alpha_2+i\beta_2)\wedge(\alpha_3+i\beta_3)+
(\alpha_1-i\beta_1)\wedge(\alpha_2-i\beta_2)\wedge(\alpha_3-i\beta_3) .
$$
\end{enumerate}
Hitchin has associated with such a nondegenerate form a tensor 
$K_\omega:N\rightarrow TN\otimes T^*N$ which is either an almost
complex structure if $\lambda(\omega)<0$ or an almost product
structure if $\lambda(\omega)>0$. Moreover, Lychagin and Roubtsov 
(1983) have defined a metric $g_\omega$ on $N$ which is compatible 
with $K_\omega$ and which has signature $(3,3)$ if $\lambda(\omega)>0$
and $(6,0)$ or $(4,2)$ if $\lambda(\omega)<0$. So, one can
associate with a nondegenerate Monge--Amp\`ere structure
$(\Omega,\omega)$ on a $6$-dimensional manifold an almost K\"ahler
structure (which is "real" if $\lambda(\omega)>0$) whose K\"ahler
form is $\Omega$. Moreover this almost K\"ahler structure is
"normalized" by the two decomposable three forms associated with
$\omega$: we use then the terminology of generalized Calabi--Yau
structure (Banos 2002).

For example, the Monge--Amp\`ere structure associated with the
"real" MAE in three variables $(x,y,z)$
$$
\hess(\phi)=1
$$
is, in the coordinates $(x,y,z,p,q,r)$, the pair
$$
\begin{cases}
\Omega=\drm x\wedge \drm p+ \drm y\wedge \drm q + \drm z\wedge \drm r&\\
\omega=\drm p\wedge \drm q\wedge \drm r -\drm x\wedge \drm y\wedge \drm z&\\
\end{cases}
$$
and the real K\"ahler structure on $T^*\mathbb{R}^3$ is
$$
\begin{cases}
g_\omega=\begin{pmatrix} 0&Id\\Id&0\end{pmatrix} ,&\\
K_\omega=\begin{pmatrix} Id&0\\0&-Id\end{pmatrix} .
\end{cases}
$$
The Monge--Amp\`ere structure associated with the special Lagrangian equation
$$
\nabla^2 \phi - \hess(\phi)=0
$$
is the pair
$$
\begin{cases}
\Omega=\drm x\wedge \drm p+ \drm y\wedge \drm q + \drm z\wedge \drm r&\\
\omega=\Im( (\drm x+i\drm p)\wedge (\drm y+i\drm q)\wedge (\drm z+i\drm r))&\\
\end{cases}
$$
and the underlying K\"ahler structure is the canonical K\"ahler
structure on $T^*\mathbb{R}^3=\mathbb{C}^3$:
$$
\begin{cases}
g_\omega=\begin{pmatrix} Id&0\\0&Id\end{pmatrix}&\\
K_\omega=\begin{pmatrix} 0&-Id\\Id&0\end{pmatrix}&\\
\end{cases}
$$
It is important to note that the geometry associated with a MAE
$\triangle_\omega=0$  of real type ($\lambda(\omega)>0$) is
essentially real but it is very similar to the classic K\"ahler
geometry. In particular, when this geometry is integrable, there
exists a potential $\Phi$ and a coordinate systems
$(x_i,p_i)_{i=1,2,3}$ on $T^*\mathbb{R}^6$ such that
$$
g_\omega=\underset{i,j}{\sum} \frac{\partial ^2 \Phi}{\partial
x_i\partial p_j} \drm x_i\cdot \drm p_j
$$
and
$$
\det\left(\frac{\partial^2 \Phi}{\partial x_j\partial
p_j}\right)=f(x)g(p).
$$
We have observed that the (elliptic) Monge--Amp\`ere equation plays an intriguing role in 
incompressible two-dimensional and the \twohd Navier-Stokes equations, and that this is 
made explicit via the introduction of the stream function. 
As we remark in \S\ref{S5}, for more general 
three-dimensional Navier-Stokes flows, a stream function can play a role as suggested in 
(\ref{vecpot1}); here we examine perhaps the simplest choice
\beq
\label{eq:Apsi} \AA = (-\psi,-\psi,-\psi), 
\eeq 
where now $\psi=\psi(x,y,z,t)$ (the minus signs are used simply to show the relationship with 
the two-dimensional case viz. $\vv = \kk\times\nabla\psi=\nabla\times (0,0,-\psi)$). Note that
(\ref{eq:Apsi}) is simply one choice that leads to some interesting features as described below. 
\textit{This choice of vector potential is, of course, very special, and merely serves to 
illustrate the possible connection with generalized Calabi--Yau structures.}

If we substitute (\ref{vp}) with (\ref{eq:Apsi}) for $\uu$ into (\ref{eq:4}) (and set $\rho = 1$), 
and use the notation $\psi_x = p, \psi_y = q, \psi_z = r$, then we find that we have another 
Monge--Amp\`ere equation that can be written as
\begin{eqnarray}
\label{eq:3d2} \omega_{3d}\equiv \nabla^2 P\drm x\wedge\drm y\wedge\drm z & + & 2\;
\drm x\wedge\drm p \wedge\drm r + 2\; \drm r\wedge\drm q\wedge\drm
z + 2\; \drm q\wedge\drm y\wedge\drm p \nonumber\\
& - & 2\; \drm y\wedge \drm q \wedge \drm r - 2\; \drm p\wedge
\drm q\wedge \drm x - 2\; \drm r\wedge \drm z\wedge \drm p
\nonumber\\
 & - & 2\;\drm x\wedge\drm q\wedge\drm r -2\; \drm p\wedge\drm
y\wedge\drm r - 2\;\drm p\wedge\drm q\wedge \drm z;\nonumber\\
&\mbox{}& \triangle_{\omega_{3d}}=0 .
\end{eqnarray}
It is worth mentioning that 
since the canonical symplectic form $\Omega=\drm x\wedge \drm p +
\drm y\wedge \drm q  + \drm z\wedge \drm r$ vanishes on the graph
of $\psi$, one can replace $\omega$ by $\omega + \theta\wedge
\Omega$, with $\theta$ any $1$-form on $T^*\mathbb{R}^3$. Following
Lychagin \& Roubtsov (1983), we choose among all these forms, the
unique one which is effective (its product with $\Omega$ is zero).
This form is
$$
\omega_0=\omega-\frac{1}{2}(\bot\omega)\wedge\Omega,
$$
with
$$
\bot{\omega}=\left(\frac{\partial}{\partial x}\wedge
\frac{\partial}{\partial p} + \frac{\partial}{\partial y}\wedge
\frac{\partial}{\partial q}+ \frac{\partial}{\partial z}\wedge
\frac{\partial}{\partial r}\right) \big{\lrcorner}\; \omega.
$$
Three invariants are associated with the effective form (\ref{eq:3d2}): the
Lychagin-Roubtsov metric $Q$, the Hitchin tensor $K$ and the
Hitchin pfaffian $\lambda$ (see Hitchin (2000) and Banos (2002)).
They are defined by
\begin{eqnarray}
\label{eq:cy1}
Q(U,V) & = & -\frac{1}{4}\bot^2\left((U\lrcorner \omega_0)\wedge
(V\lrcorner \omega_0)\right)\,,\\
\label{eq:cy2}
 K_{ij}& = &\Omega_{ik} Q_{kj}\,,\\
\label{eq:cy3}
 \lambda & = & \frac{1}{6} \tr(K^2)\,.\\
\end{eqnarray}
When $\lambda<0$, the tensor $J=\frac{1}{\sqrt{|\lambda|}}K$ is an
almost complex structure and the pair $(Q,J)$ describes a
Calabi--Yau geometry.
We obtain here $\lambda=128\nabla^2 P$: our equation  is associated
with a complex geometry if and only if 
\bel{3dlap}
\nabla^2 P < 0\,.
\ee
The Lychagin-Roubtsov metric is
\begin{equation}{\label{metric1}}
Q=16\begin{pmatrix} \alpha^2&\alpha^2&\alpha^2&0&0&0\\
\alpha^2&\alpha^2&-\alpha^2&0&0&0\\
\alpha^2&-\alpha^2&\alpha^2&0&0&0\\
0&0&0&0&1&1\\
0&0&0&1&0&-1\\
0&0&0&1&-1&0\\
\end{pmatrix}
\end{equation}
and the Hitchin complex structure is
\begin{equation}{\label{CS1}}
J=\frac{1}{\sqrt{2}\alpha}\begin{pmatrix} 0&0&0&0&1&1\\
0&0&0&1&0&-1\\
0&0&0&1&-1&0\\
-\alpha^2&-\alpha^2&-\alpha^2&0&0&0\\
-\alpha^2&-\alpha^2&\alpha^2&0&0&0\\
-\alpha^2&\alpha^2&-\alpha^2&0&0&0\\
\end{pmatrix}\,,
\end{equation}
with $\nabla^2 P = -4\alpha^2$. Once again, this complex structure
is integrable in the special case when $\nabla^2 P$ is constant. 

\begin{center}
{\sc References}
\end{center}
\noindent
Banos, B. (2002) Nondegenerate Monge--Amp\`ere structures in
dimension 6, {\em Lett. Math. Phys.}, {\bf 62} 1-15.

\noindent
Burgers, J. M. (1948) A mathematical model illustrating the theory
of turbulence. {\em Adv. Appl. Math.}, {\bf 1}, 171-199.

\noindent
Constantin, P. (2000) The Euler equations and non-local conservative 
Riccati equations, \textit{Internat. Math. Res. Notices (IMRN)}, 
\textbf{9}, 455-65.

\noindent
Doering, C. R.  \& Gibbon, J. D. (1995) {\em Applied
analysis of the Navier-Stokes equations}. Cambridge: University
Press, 217pp.

\noindent
Gibbon, J. D. (2002) A quaternionic structure in the
three-dimensional Euler and ideal magneto-hydrodynamics equations.
\textit{Physica D}, {\bf 166}, 17-28.

\noindent
Gibbon, J. D., Fokas, A. \& Doering, C.R. (1999) Dynamically
stretched vortices as solutions of the Navier-Stokes equations. 
\textit{Physica D}, \textbf{132}, 497-510.

\noindent
Hitchin, N. J. (2000) The geometry of three-forms in six dimensions. {\em J. Diff. Geom.}, {\bf 55}, 547-76.

\noindent
Hoskins, B. J. \& Bretherton, F. P. (1972) Atmospheric frontogensis models: mathematical formulation and solutions. {\em J. Atmos. Sci.}, {\bf 29}, 11-37.

\noindent
Kerr, R. M. (1993) Evidence for a singularity of the three-dimensional 
incompressible Euler equations. {\em Phys. Fluids} A \textbf{6}, 1725.

\noindent
Kerr, R.M. (2005) Vorticity and scaling of collapsing Euler vortices, 
\textit{Phys. Fluids}, \textbf{17}, 075103-114.

 \noindent
Lundgren, T. (1982) Strained spiral vortex model for turbulent
fine structure, \textit{Phys. Fluids}, \textbf{25}, 2193-2203.

\noindent
Lychagin, V. V. (1979) Nonlinear differential equations and contact geometry. 
{\em Usp\`ekhi Mat. Nauk.}, {\bf 34}, 137-65.

\noindent
Lychagin, V. V. \& Roubtsov, V. N. (1983) Local
classifications of Monge--Amp\`ere equations. {\em Dokl. Bielor.
Acad. Sci.}, {\bf 27} 396-98.

\noindent
Lychagin, V. V., Roubtsov, V. N. \& Chekalov, I. V. (1993) A classification 
of Monge--Amp\`ere equations, {\em Ann. Sci. Ec. Norm. Sup.}, {\bf 26}, 281-308.

\noindent
McIntyre, M. E. \& Roulstone, I. (2002) Are there
higher-accuracy analogues of semi-geostrophic theory? In {\em
Large-scale atmosphere---ocean dynamics, Vol. II: Geometric
methods and models}. (Eds.) J. Norbury and I. Roulstone; Cambridge: University Press.

\noindent
Majda, A. J. (1986) Vorticity and the mathematical theory of
incompressible fluid flow, {\em Comm. Pure and Appl. Math.}, {\bf 39}, 187-220.

\noindent
Majda, A. J. \& Bertozzi, A. (2002) \textit{Vorticity and incompressible 
flow}, Cambridge: University Press.

\noindent
Moffatt, H. K., Kida, S. \& Ohkitani, K. (1994) Stretched vortices --
the sinews of turbulence; large-Reynolds-number asymptotics. {\em J. Fluid Mech.},
{\bf 259}, 241-264.

\noindent
Ohkitani, K. \& Gibbon, J. G. (2000) Numerical study of singularity 
formation in a class of Euler and Navier-Stokes flows, {\em Physics of Fluids}, 
\textbf{12}, 3181-3194.

\noindent
Pullin, D. I. \& Saffman, P. G. (1998) Vortex dynamics, {\em Ann.
Rev. Fluid Mech.}, {\bf  30}, 31-51.

\noindent
Roubtsov, V. N. \& Roulstone, I. (1997) Examples of
quaternionic and K\"ahler structures in Hamiltonian models of
nearly geostrophic flow, {\em J. Phys. A}, {\bf 30}, L63-L68.

\noindent
Roubtsov, V. N. \& Roulstone, I. (2001) Holomorphic structures in hydrodynamical 
models of nearly geostrophic flow. {\em Proc. R. Soc. Lond.}, A {\bf 457}, 1519-1531.

\noindent
Roulstone, I \& Sewell, M. J. (1997) The mathematical structure of theories of semi-geostrophic type. {\em Phil. Trans. R. Soc. Lond.}, {\bf A355}, 2489-2517.

\noindent
Saffman, P. G. (1993) \textit{Vortex Dynamics}, Cambridge: University Press.

\noindent
Stuart, J. T. (1987) Nonlinear Euler partial differential equations:
singularities in their solution, In {\em Proceedings in honor of C. C. Lin}
(edited by D. J. Benney, Chi Yuan and F. H. Shu), pp 81-95, World Scientific,
Singapore.

\noindent
Stuart, J. T. (1991) The Lagrangian picture of fluid motion and its 
implication for flow structures, {\em IMA J. Appl. Math.}, {\bf 46}, 
147.

\noindent
Vincent, A. \& Meneguzzi, M. (1994) The dynamics of vorticity tubes of
homogeneous turbulence, {\em J. Fluid Mech.}, {\bf 225}, 245-254.

\end{document}